\documentclass[doublecol]{epl2}

\usepackage{amssymb,amsmath,latexsym}

\usepackage{epsfig,psfrag,subfigure}
\usepackage{graphicx,psfrag,subfigure}
\usepackage{color}
\newcommand\beq{\begin{equation}}
\newcommand\eeq{\end{equation}}
\newcommand\bea{\begin{eqnarray}}
\newcommand\eea{\end{eqnarray}}
\newcommand\al{\alpha}
\newcommand\be{\beta}

\newcommand\de{\delta}
\newcommand\De{\Delta}
\newcommand\si{\sigma}
\newcommand\dg{\dagger}
\newcommand\la{\langle}
\newcommand\ra{\rangle}

\newcommand\tf{\tilde f}

\newcommand\ig{\includegraphics}
\newcommand\bib{\bibitem}

\title{Quenching across quantum critical points: role of topological patterns}

\author{Diptiman Sen\inst{1} \and Smitha Vishveshwara\inst{2}}

\institute{\inst{1} {Centre for High Energy Physics, Indian Institute of 
Science, Bangalore 560012, India} \\
\inst{2} {Department of Physics, University of Illinois at Urbana-Champaign, 
1110 W. Green St, Urbana, IL 61801, USA}}

\date{\today}

\abstract
{We introduce a one-dimensional version of the Kitaev model consisting 
of spins on a two-legged ladder and characterized by 
$Z_2$ invariants on the plaquettes of the ladder. We map the model to a 
fermionic system and identify the topological sectors associated with 
different $Z_2$ patterns in terms of fermion occupation numbers. Within 
these different sectors, we investigate the effect of a linear quench across 
a quantum critical point. We study the dominant behavior of the 
system by employing a Landau-Zener-type analysis of the effective Hamiltonian 
in the low-energy subspace for which the effective quenching can sometimes 
be non-linear. We show that the quenching leads to a residual energy which 
scales as a power of the quenching rate, and that the power depends on the 
topological sectors and their symmetry properties in a non-trivial way. This 
behavior is consistent with the general theory of quantum quenching, but with 
the correlation length exponent $\nu$ being different in different sectors.}


\begin{document}

\maketitle

Of late, two different concepts have instigated a surge of active research 
in quantum many-body phenomena - the notion of topological order 
\cite{nayaketal08,levin05,kitaev06, sarma06} and quenching across quantum 
critical points (QCPs) \cite{kibble76,zurek05,levitov06,mondal,dziarmaga09}.
The former, described by global invariants, has been keenly studied for its 
fundamental significance, realizations in physical systems, associated phase 
transitions and potential applications to quantum computation. The latter has 
offered a mine of valuable information on the nature of the QCP in question, 
for instance, in the scaling behaviors of the defect density and the residual 
energy upon quenching. These quantities are governed by a quantum version of 
the Kibble-Zurek mechanism and in $d$ dimensions, they scale with the 
quenching rate $1/\tau$ as $1/\tau^{d\nu/(z\nu +1)}$, where $\nu$ and $z$ are 
respectively the correlation length and dynamical critical exponents of the
QCP. Post-quench dynamics is thus effectual in extracting 
a combination of these exponents defined by the growth of the correlation 
length as $|\lambda - \lambda_c|^{-\nu}$ and the relaxation time as 
$|\lambda - \lambda_c|^{-z\nu}$, as a parameter $\lambda$ in the Hamiltonian 
of the system approaches a critical value $\lambda_c$ \cite{kibble76,zurek05}.
In this Letter, we demonstrate that a marriage between these two concepts 
makes for a synergistic union. Specifically, we show that invariants in a 
system having topological order can impose dramatic constraints
on statics and dynamics, giving rise to qualitatively different post-quench 
behaviors; hence, quenching can act as a powerful means of discerning 
patterns and symmetries in the topological invariants of a system.

Our focus here is on models sharing traits of the celebrated Kitaev model 
\cite{kitaev06}, namely spin-1/2 lattice systems with an exponentially 
large number of configurations distinguished by conserved $Z_2$ quantum 
numbers, which we refer to as topological sectors. Detailed studies 
characterizing the various topological sectors and 
distinguishing their physical effects are sparse; here we address these 
issues in the context of quenching in a simple two-legged ladder model which 
captures salient features of the two-dimensional parent Kitaev system. 
We explicitly characterize the $Z_2$ patterns based on their periodicity and, 
within these different sectors, study quenching at zero temperature through a
QCP. We find that the residual energy, the difference between the post-quench
energy expectation value and the true ground state energy, respects the 
$d\nu/(z \nu + 1)$ power-law scaling form with respect to the quench rate. 
While the dynamical exponent takes the value $z=1$, the value of $\nu$, and 
therefore the power-law, depends on the topological sector being probed. For 
sectors endowed with higher symmetry in their periodicity, $\nu$ deviates 
from unity and the power-law is no longer of the $1/\tau^{1/2}$ form typical 
of many one-dimensional systems \cite{kibble76,zurek05}. We thus 
theoretically demonstrate, using a simple model, the effect of topology on 
quantum critical behavior and the rich resultant physics in quench dynamics.

\begin{figure}[t] \ig[width=3in]{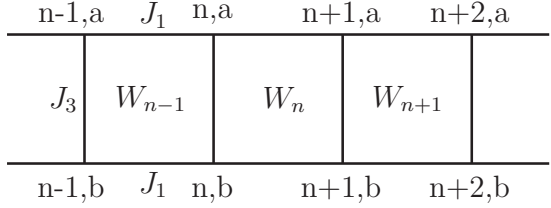}
\caption[]{Picture of the two-legged Kitaev ladder showing the couplings
$J_i$ and the $Z_2$-valued operators $W_n$.} \label{fig:ladder} \end{figure}

The quasi-one-dimensional Kitaev model that we consider here (see Fig. 
\ref{fig:ladder}) consists of a two-legged ladder with a spin-$1/2$ at each 
site. We label the sites as $(n,a)$ and $(n,b)$, where $a,b$ denote the upper 
and lower legs respectively. The Hamiltonian for the system is given by
\beq H = \sum_n [J_1 (\si_{n,a}^x \si_{n+1,a}^y + \si_{n,b}^x \si_{n+1,b}^y) 
+ J_3 \si_{n,a}^z \si_{n,b}^z], \label{ham1} \eeq
where $\si^\al_{n,a/b}$ denote the Pauli matrices; the couplings considered 
here are analogous to those associated with a strip of the honeycomb lattice 
in the parent Kitaev model. (We set $\hbar =1$.) We assume $J_1>0$; negative 
$J_1$ values can be gauged away by performing $\pi$-rotations of appropriate 
spin components. 
As discussed in Ref. \cite{feng07}, the system exhibits a rich 
phase diagram as a function of the couplings $J_1$ and $J_3$.
Each plaquette of the ladder hosts a topological invariant and together 
these invariants characterize different topological sectors. For a given 
plaquette (see Fig. \ref{fig:ladder}), we can define a Hermitian operator 
$W_n = \si_{n,a}^y \si_{n,b}^y \si_{n+1,a}^x \si_{n+1,b}^x$. One can show 
that the $W_n$'s commute with each other and with the Hamiltonian, and that 
$W_n^2 = 1$. The invariant eigenvalues, $W_n=\pm 1$, provide a set of $Z_2$ 
quantum numbers and, for a system with $N$ plaquettes, yield $2^N$ distinct 
sectors corresponding to different sets of these numbers. 

As in previous studies \cite{feng07,nussinov08}, we map the system to that of 
two sets of spinless fermions, one probing the dynamics and the other 
associated with the Kitaev model's hallmark Ising anyons \cite{nayaketal08}. 
The mapping involves a Jordan-Wigner transformation which takes the 
form $a_n = S_{n,a} ~\si^y_{n,a}$, $c_n = S_{n,a} ~\si^x_{n,a}$, 
$b_n = S_{n,b} ~\si^x_{n,b}$, $d_n = S_{n,b} ~\si^y_{n,b}$, 
where $S_{n,a/b}$ denote a string of $\si^z_{n,a/b}$'s defined 
as follows. Assuming that each leg contains $N$ sites with open (rather 
than periodic) boundary conditions, $S_{n,a} =\prod_{m=1}^{n-1} ~
\si^z_{m,a}$ weaves from the left to the right along the top leg in 
Fig. \ref{fig:ladder} to the $(n-1)$-th site, and $S_{n,b} = \prod_{m=1}^N ~
\si^z_{m,a} ~\prod_{m=n+1}^N ~\si^z_{m,b}$ weaves completely along the 
top leg left to right and then along the bottom leg from right to left to 
the $(n+1)$-th site. The transformed Hermitian operators are Majorana fermion
operators satisfying relations such as $a_n^2 = 1$, $\{ a_m,a_n \} = 2 
\de_{mn}$, $\{ a_m,b_n \} = 0$, and so on. The resultant Hamiltonian takes 
the form $H = \sum_n [-i J_1 (a_n a_{n+1} + b_n b_{n+1}) + i (-1)^n J_3 a_n 
b_n \prod_{m=n}^N W_m]$, where $W_n = - (i c_n d_n) (i c_{n+1} d_{n+1})$. 
Thus, the Hamiltonian is quadratic in the $a-b$ fermions and dependent on 
the $Z_2$ values of the topological invariants.

Next, the Majorana fermions can be combined pairwise to form Dirac fermions, 
providing an understanding of the system in terms of Dirac fermion occupation
numbers. The natural combinations, which pair partners on each rung of the 
ladder, are $f_n = (i^n/2) (a_n + i b_n)$ and $g_n = (1/2) (c_n + i d_n)$;
these satisfy $\{ f_m,f_n^\dg \} = \{ g_m,g_n^\dg \} = 
\de_{m,n}$. The topological invariants can be expressed in terms of occupation
numbers of the $g$-fermions ($g_n^\dg g_n=0,1$) as $W_n = -s_n s_{n+1}$ where 
$s_n = ic_n d_n = 2g_n^\dg g_n - 1$. The value of $W$ on the $n$-th 
plaquette is $+1$ if its bordering $g$-fermions have opposite occupation 
numbers and $-1$ if they are the same. The $s_n$'s themselves are non-local 
in terms of the $W_n$'s: $s_n = \prod_{m=n}^\infty W_m$. Specifying the 
values $s_n=\pm 1$ completely fixes those of the $W_n$'s. For instance, 
alternating values of $s_n$ along the ladder gives all $W_n = +1$, a constant 
value of $s_n$ gives all $W_n = -1$, and a domain wall in $s_n$ is equivalent 
to embedding a $W_n=+1$ at the wall amidst all other $W_n$'s being $-1$. 
[Changing the values of $W_n$ and thus the topological sector can be 
achieved by the appropriate local action of operators such as $\si_{n,a}^y 
\si_{n+1,a}^x$ and $ \si_{n,b}^y \si_{n+1,b}^x$, which can alter the 
occupation numbers of the $g$-fermions on any given site]. 
In the thermodynamic limit, the Hamiltonian in 
Eq. (\ref{ham1}) takes the local form
\beq H = \sum_n [-2 J_1 (f_{n+1}^\dg f_n + f_n^\dg f_{n+1}) + J_3 s_n 
(2f_n^\dg f_n - 1)], \label{ham4} \eeq
namely a tight-binding system for the $f$-fermions whose on-site chemical 
potential is determined by the $g$-fermion occupation number. The Hilbert 
space of $2^{2N}$ associated with the $2N$ spins on a $N$-plaquette ladder is 
thus split into two Hilbert spaces, each of dimension $2^N$,
corresponding to those of the `dual' $f$ and $g$ fermions (whose roles can 
be exchanged by swapping the $x$ and $y$ spin components.) 
The $f$-fermions probe the dynamics of the system, while the $g$-fermions, 
whose Hilbert space dimension reflects the $SU(2)_2$ structure associated 
with Ising anyons \cite{nayaketal08}, determine its topology.

The topological sectors for this model can be identified by the patterns 
exhibited by the $s_n$'s. For patterns of $s_n$ having a period $p$, the 
Hamiltonian of Eq. (\ref{ham4}) splits up into decoupled sub-systems involving
$p$ momentum states with values $k+2\pi q/p$, where $q$ is an integer 
satisfying $0 \le q \le p-1$ and $k$ lies in the range $-\pi < k \le -\pi + 2
\pi/p$. To determine the filling of the $f$-fermions for the lowest energy 
state 
within a given sector, consider the limit $J_1=0$ and $J_3 < 0$. Here energy 
minimization requires that the occupation number $f_n^\dg f_n$ at site $n$ be 
equal to $(1+s_n)/2$; the gap to the first excited state is given by $2|J_3|$.
A periodicity having $m$ of the $s_n$'s be $+1$ and $p-m$ be $-1$ thus 
has $m$ sites occupied and the others unoccupied within a period; the filling 
is $m/p$. Turning on $J_1$ does not change the particle number; hence the 
filling remains constant in the lowest energy state. As can be shown 
perturbatively around the $J_1=0$ limit, a small value of $J_1$ changes the 
ground state energy, to second-order in $J_1$, by $\De E_0 = -(J_1^2/J_3)
\sum_n (1-s_n s_{n+1})$. Hence, the ground state sector in which the energy 
of the lowest-lying state is the minimum amidst all the sectors is the one 
in which $s_n = (-1)^n$ and $W_n = 1$ for all $n$. 

Having characterized the system by the different topological sectors and 
their eigenstate structures, 
we now investigate the effects of quenching within different topological 
sectors. We assume that the system is initially placed in a particular sector;
a key feature of a topological sector is that the system will remain in that 
sector at all times if there are no perturbations which can dynamically change
the topological invariants. We initialize the system in its lowest energy 
state at $J_3 = -\infty$; at that point, the energy gap between the ground 
state and
the excited states is infinitely large. We then vary $J_3$ linearly in time as
$J_3 (t) = J_1 t/\tau$; this quench tunes the system through a QCP at $J_3 =0$
as discussed below. We quantify the effect of the quench via the residual 
energy per site attained at the final time, which is defined as 
\beq E_r ~=~ \lim_{t,N \to \infty} ~\frac{\la H \ra_f - E_0}{N|J_3(t)|}, 
\label{er1} \eeq
where $\la H \ra_f$ denotes the expectation value of the Hamiltonian in the 
final state reached, and $E_0$ is the ground state energy at $t \to \infty$. 
We study the dependence of $E_r$ on $\tau$ for $J_1 \tau \gg 1$; this must 
vanish in the adiabatic limit $\tau \rightarrow \infty$. We find that 
$E_r$ shows different qualitative trends depending on the manner in which the 
topologically sensitive $J_3$ term in Eq. (\ref{ham4}) couples different 
momentum modes. We discuss three distinct classes of behavior: (i) the trivial
case of decoupled modes, (ii) a Landau-Zener type $\tau^{-1/2}$ dependence of 
$E_r$ due to direct coupling between the low-energy modes, and (iii) the most 
interesting case of a non-trivial power-law due to an indirect coupling. 

Let us represent the $f$-fermions in the momentum 
basis as $\tf_k =\frac{1}{\sqrt N} \sum_n f_n e^{-ikn}$, where $-\pi < k \le 
\pi$. In this basis, the sector having all $W_n =-1$, i.e., all $s_n = 1$ for 
all $n$, can be seen to behave trivially under quenching. The modes having 
different momenta do not mix for any $J_3$ in the Hamiltonian of Eq.
(\ref{ham4}). Thus, if one begins in the lowest energy state for $J_3 = -
\infty$, i.e., with $f_n^\dg f_n = 1$, one remains in that state as $J_3$ 
changes to $\infty$ where it becomes the highest excited state. Hence, the 
residual energy $E_r$ is equal to 2 independently of the value of $\tau$.

The ground state sector having all $W_n = 1$ offers an instance of direct 
coupling between low-energy modes. Assuming that $s_n = (-1)^n$, the 
Hamiltonian decouples into sub-systems having pairs of momenta $k$ and 
$k+\pi$ as
\beq H_2 ~=~ \sum_k ~\vec{\tf}_{k2}^\dg ~\left( \begin{array}{cc}
4J_1 \cos k & 2J_3 \\
2J_3 & - 4J_1 \cos k \end{array} \right) ~\vec{\tf}_{k2}, \label{ham5} \eeq
where $\vec{\tf}_{k2} = (\tf_{k+\pi},\tf_k)^T$ represents the momentum mode 
annihilation operators ($T$ denotes the transpose), 
and $-\pi < k \le 0$. The corresponding eigenenergies are 
$E_{k\pm} = \pm \sqrt{4J_3^2 + 16 J_1^2 \cos^2 k}$. For $J_3=0$, the energies 
vanish at $k = - \pi/2$, indicative of a quantum critical point (QCP); in 
fact, a QCP having gapless modes at $k_c = \pm \pi/2$ occurs at $J_3 = 0$ in 
any topological sector with a half-filled ground state. Here, the dynamical 
critical exponent is $z=1$ and the correlation length exponent is $\nu = 1$ 
since the energy vanishes as $|k-k_c|$ at the QCP at $J_3=0$ and as $|J_3|$ 
at $k = k_c$.

For the quench $J_3= J_1 t/\tau$, a $\pi/2$ unitary rotation interchanges the 
diagonal and off-diagonal terms in the matrix appearing in Eq. (\ref{ham5}), 
exactly mapping the quench to the well-known Landau-Zener problem 
\cite{lz,zurek05,levitov06}. The probability $p_k$ of ending in 
an excited state at $t \to \infty$ and the net residual energy, for which each
sub-system contributes $4p_k$, are therefore given by
\beq p_k ~=~ \exp [- 8 \pi \tau J_1 \cos^2 k], \quad E_r ~=~ \int_{-\pi}^0 ~
\frac{dk}{2\pi} ~4 p_k. \label{er2} \eeq
The probability $p_k$ is largest for the low-energy modes near $k_c = - \pi/2$
since this is where the gap between the two states vanishes for $J_3 = 0$. In 
the limit $\tau J_1 \to \infty$, the residual energy is dominated by this 
low-energy regime; expanding around $k_c$
yields a Gaussian integral and the power-law form $E_r \sim 1/ \tau^{1/2}$. 
This scaling is consistent with the values of the critical exponents given 
above and in keeping with quenches through QCPs in many one-dimensional 
systems \cite{kibble76,zurek05}. 

We now turn to an instance of a topological sector yielding a completely 
different quench power-law of $1/\tau^{2/3}$ due to a higher symmetry in its 
periodic structure, as we show by heuristic arguments and numerics. The 
instance is of $W_n = (-1)^n$, or equivalently, the signs of $s_n$ forming 
the half-filling period 4 pattern ($++--$). The pattern can be expressed
as $s_n=- \sqrt{2} \cos (\pi n/2 + \pi/4)$. The resulting Hamiltonian of 
Eq. (\ref{ham4}) decouples into sub-systems involving four momenta with 
annihilation operators $\vec{\tf}_{k4}= (\tf_{k+3\pi/2},\tf_{k+\pi},
\tf_{k+\pi/2}, \tf_k)^T$, where $k$ lies in the range $-\pi < k \le -\pi/2$. 
Explicitly, $H_4 = \sum_k \vec{f}_{k4}^\dg H_{k4} \vec{f}_{k4}$, where 
$H_{k4} = M_k+N_k$ with
\bea M_k &=& -J_3\sqrt{2} ~\left[ \left( \begin{array}{cc}
0 & \be \\
\be^* & 0 \end{array} \right) \otimes I + \left( \begin{array}{cc}
0 & \be^* \\
\be & 0 \end{array} \right) \otimes \mu^x \right], \label{ham6a} \\
N_k &=& 4J_1~\left( \begin{array}{cc}
- \sin k & 0 \\
0 & \cos k \end{array} \right)~\otimes~\mu^z. \label{ham6b} \eea
Here $\be = e^{i\pi/4}$, $\mu^\alpha$ denote Pauli matrices and $I$ is the 
identity matrix. The higher symmetry of this sector is reflected by the fact 
that while the four momentum states cannot be decoupled into pairs, they are 
not all directly coupled to one another. The eigenenergies of $H_{k4}$ come 
in pairs $\pm E_{k4}$, as can be derived from the symmetry property $UH_{k4} 
U^\dg = - H_{k4}$, where $U=\mu^z\otimes \mu^x$. The QCP associated with this 
system occurs, as reflected in Eqs. (\ref{ham6a}-\ref{ham6b}), at $J_3=0$ and 
$k$ close to $-\pi$ and $-\pi/2$. For a quench $J_3 = J_1 t/\tau$, the 
dominant contribution to the residual energy comes from particle-hole pairs 
closest to the Fermi level at zero energy for which the cost of 
quenching into an excited state is low. For instance, for momenta $k 
\gtrapprox -\pi$, the relevant pairs closest to the Fermi energy at small 
$J_3$ are $|1, 0\rangle$ and $|0, 1\rangle$, where $n_p$ and $n_0$ in state 
$|n_p, n_0\rangle$ denote $f$-fermion occupation numbers of momenta $k$ and $k 
+ \pi$, respectively. These states are not directly coupled by the 
Hamiltonian $H_{k4}$ and only mix via the higher energy states at $k + \pi/2$ 
and $k + 3\pi/2$. To second-order perturbation in $J_3$, this coupling is of 
order $J_3^2/J_1$; the effective Hamiltonian for the two-level system, in 
the appropriate basis, takes the form
\beq H_{k4,eff} ~=~ J_1 ~\left( \begin{array}{cc}
t^2/\tau^2 & 4 k \\
4 k & - t^2/\tau^2 \end{array} \right), \label{ham8} \eeq
where we have expanded $-\sin (k - \pi) \simeq k$, and have used the 
time-dependent form of $J_3= J_1t/\tau$. A similar form applies to modes 
at $k \lessapprox - \pi/2$. Arguments similar to the previous case show 
that here $z=1$ and $\nu=2$.

\begin{figure}[t] \ig[width=1.6in]{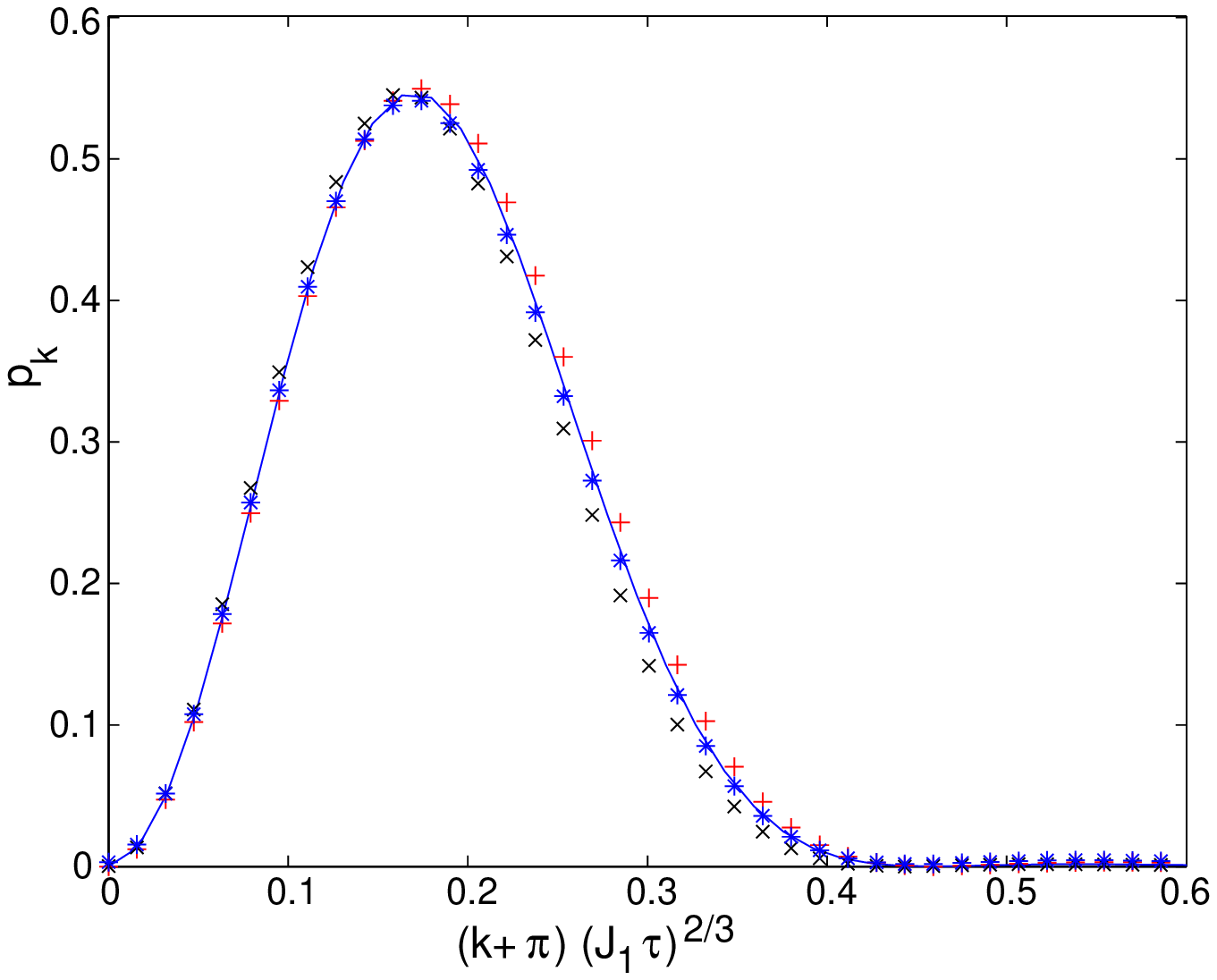}\ig[width=1.6in]{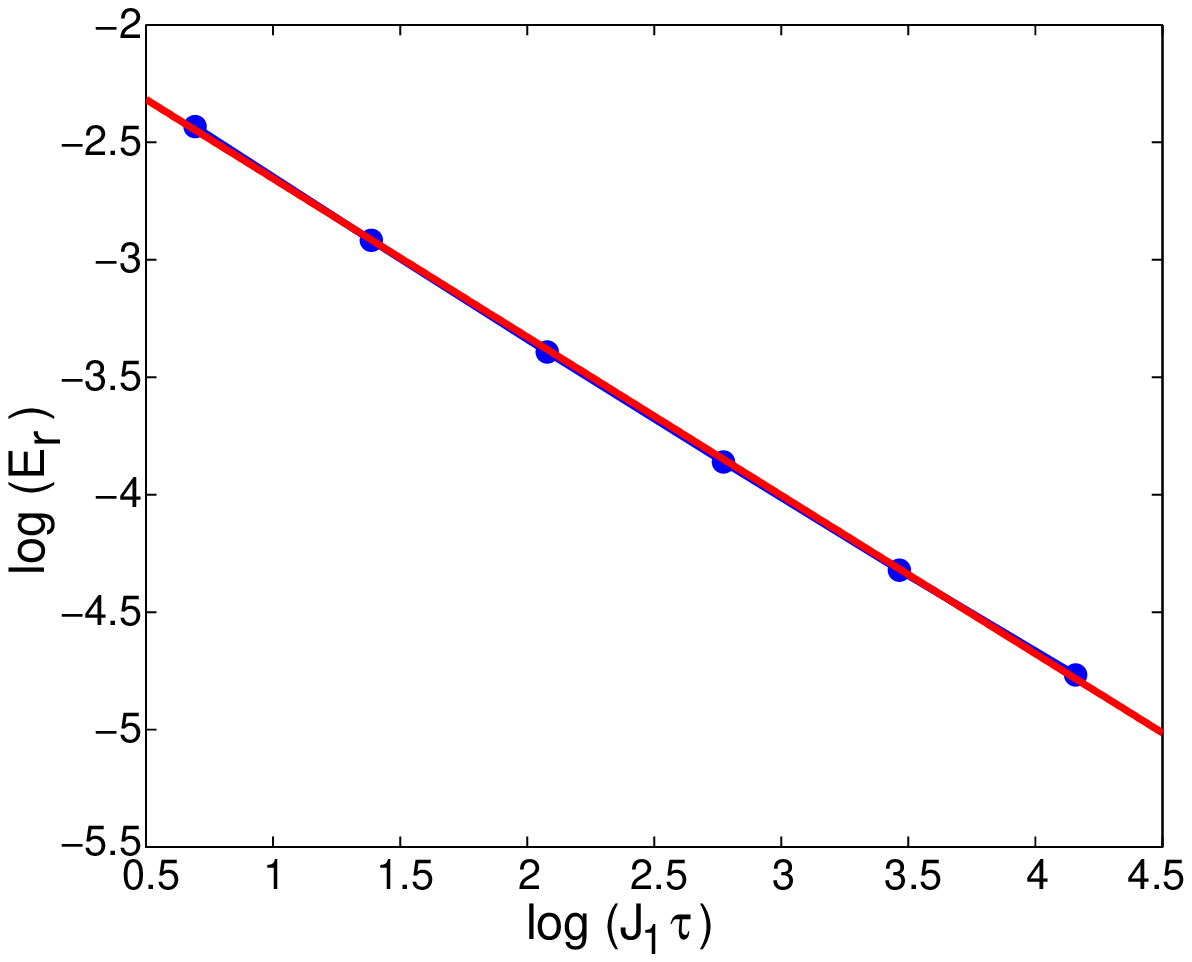}
\caption[]{Left panel: Plot of $p_k$ versus $(k+\pi)(J_1 \tau)^{2/3}$ for $k 
\gtrapprox - \pi$, and $J_1 \tau=2$ (red +), 8 (black x) and 32 (blue $*$). 
Right panel: Logarithmic plot of $E_r$ versus $J_1 \tau$ in the sector with 
$W_n = (-1)^n$. A linear fit gives $E_r = 0.14/(J_1 \tau)^{0.67}$ which is 
close to a $-2/3$ power-law.} \label{fig:ervstau} \end{figure}

The effective quench in Eq. (\ref{ham8}) is quadratic in time. 
Unlike the linear quenching problem, no exact solution is known for the 
excitation probability $p_k$ for the Schr\"odinger equation
$i d\psi_k /dt = H_{k4,eff} \psi_k$ given by Eq. (\ref{ham8}). However, 
we can invoke scaling arguments to find the power-law dependence of $p_k$ on 
$\tau$. Rescaling time as $t'= t /\tau^{2/3}$ results in the excitation 
probability being governed by the parameter $k \tau^{2/3}$. If $k \tau^{2/3} 
\lesssim 1$, the corresponding modes have a significant weight for occupying 
the excited state as $t' \to \infty$, while the modes with $k \tau^{2/3} 
\gg 1$ remain in the ground state. Integrating $p_k$ around the low-energy 
modes to obtain the residual energy thus results in the scaling $E_r \sim 
1/\tau^{2/3}$. We ascertain this behavior by numerically solving the 
time-dependent Schr\"odinger equation given by the full-fledged
Hamiltonian in Eqs.  (\ref{ham6a}-\ref{ham6b}) from $t = -\infty$ to $\infty$ 
with $J_3 = J_1 t/\tau$. We start in the lowest energy state at $J_3 =-\infty$ 
consisting of the two occupied states $(1,-1,1,-1)/2$ and $(1,i,-1,-i)/2$, and 
time evolve to obtain the probability $p_k$ for occupying the excited states.
The left panel of Fig. \ref{fig:ervstau} shows the resultant $p_k$ versus
the scaled variable $(k+\pi)(J_1\tau)^{2/3}$ for $k \gtrapprox - \pi$; 
the curve is independent of $\tau$ for $J_1 \tau \gg 1$. The probability 
$p_k$ near $k \lessapprox - \pi/2$ is related to this curve by mirror 
symmetry. Analogous to Eq. (\ref{er2}), integrating $2p_k/\pi$ over $k$ 
from $-\pi$ to $-\pi/2$ 
yields the residual energy $E_r$. The right panel of Fig. \ref{fig:ervstau} 
shows a plot of $\log(E_r)$ versus $\log (J_1 \tau)$. The clean linear fit with
a slope close to $-2/3$ confirms our predicted $1/\tau^{2/3}$ scaling form.

This unusual power-law scaling is generic to a series of topological sectors
that are at half-filling, i.e., those that in a period $2m$ have half the 
$s_n$'s take on the value $+1$. Of the associated $2m$ momentum modes, the 
states closest to $k = \pm \pi/2$ and thus to the Fermi energy dominate the 
quench. As in the example above, the quench through the QCP at $J_3=0$ shows 
unusual scaling if these states are not directly coupled. This corresponds to 
the $J_3$ term in the Hamiltonian having no matrix element 
between the low-energy modes or, equivalently,
having $A_m$ vanish in the Fourier expansion $A_l = \sum_{j=1}^{2m} s_j 
e^{i\pi jl/m}$. Clearly the number of such cases increases with increasing 
$m$. For instance, such examples of period $8$ which also yield 
the $1/\tau^{2/3}$ power-law by way of low-energy modes being coupled via 
one intermediate high energy state are the sets ($++++----$), ($+++--+--$), 
($++-++---$), ($++-+--+-$), and cyclic permutations thereof. However, in 
cases of even higher symmetry where the low-energy modes are only coupled via 
a path involving $q$ intermediate high energy states, the effective coupling 
is of the form $J_1 (t/\tau)^{q+1}$ for the linear quench $J_3=J_1t/\tau$. 
In principle, the generalized scaling argument presented above would 
predict that the residual energy would have the scaling form $E_r \sim 
\tau^{-(q+1)/(q+2)}$ corresponding to $z=1$ and $\nu = q+1$, but the analysis
would depend on the specific topological sector and dominant couplings
close to the QCP.


In conclusion, we have demonstrated that in certain systems, topological 
order can play a major role in determining the universality class and 
behavior close to QCPs. Topology can greatly constrain dynamics and yield 
unusual power-law scaling in quenches across QCPs which can distinguish 
different orderings based on symmetry properties. The one-dimensional example 
showcased here ought to be realizable using recent schemes proposed in cold 
atomic systems \cite{coldatoms} and the associated topological sectors ought
to be accessed using local operations. 
We expect that 
topology-driven scaling forms would be manifest in several observables such as
dynamic spin-spin correlations, and in two-dimensional generalizations of the 
Kitaev model. Finally, due to a finite quantum critical regime in parameter
space and the robustness of topological sectors against a large class of 
perturbations, we expect the features discussed here to persist at 
temperatures much smaller than the energy gap at the initial time. At higher
temperatures, the interplay between thermal and quantum fluctuations on quench
dynamics \cite{patane} would require further investigation in our model.

We thank K. Sengupta, S. Deng and S. Trebst for illuminating discussions. We 
gratefully acknowledge the support of DST, India under Project No. 
SR/S2/CMP-27/2006 (DS), the NSF under the grant DMR 06-44022 CAR (SV) and 
the CAS fellowship at UIUC (SV).

\end{document}